\documentclass{sig-alternate}
\newdef{definition}{Operation}

\begin{document}
%
\conferenceinfo{SIGKDD}{'03, August 24-27, 2003, Washington, DC, USA}
\CopyrightYear{2003} 
\crdata{1-58113-737-0/03/0008}  

\title{EqRank: A Self-Consistent Equivalence Relation on Graph Vertexes}

%
%

\numberofauthors{2}
%

\author{
%
\alignauthor Grigorii Pivovarov\\
       \affaddr{Institute for Nuclear Research}\\
       \affaddr{60th October Anniversary Prospect 7a}\\
       \affaddr{Moscow, 117312 Russia}\\
       \email{gbpivo@ms2.inr.ac.ru}
\alignauthor Sergei Trunov\\
       \affaddr{SEUS}\\
       \affaddr{Bol'shoi Trekhsvyatitel'skii per. 2}\\
       \affaddr{Moscow, 109028 Russia}\\
       \email{trunov@msk.seus.ru} }
\date{3 July 2003}
\maketitle
\begin{abstract}
A new method of hierarchical clustering of graph vertexes is suggested.
In the method, the graph partition is determined
with an equivalence relation satisfying a recursive definition 
stating that vertexes are equivalent if the vertexes they point to
(or vertexes pointing to them) are equivalent. Iterative 
application of the partitioning yields a hierarchical clustering of graph
vertexes. 
The method is applied to the citation graph of \texttt{hep-th}.
The outcome is a two-level classification scheme for 
the subject field presented in \texttt{hep-th}, and 
indexing of the papers from \texttt{hep-th} in this scheme. 
A number of tests show that 
the classification obtained is adequate.
\end{abstract}




\section{Introduction}
With the advent of the Internet, scientific literature comes closer to 
realizing the notion of {\it Knowledge Network} \cite{ginsparg:creating}, 
with the
papers as the information units, and the references to other papers
as the links of the 
network. An important feature in the organization of scientific knowledge
is its representation as a hierarchy of developing and transforming 
scientific themes. The search for algorithms that
would be able to reveal this hidden hierarchy analyzing the 
network structure had been initiated in the seventies 
\cite{small:structure, garfield:system}, and continued until now 
\cite{popescul:clustering, flake:graph, eckmann:curvature}. 
Most of the present day clustering algorithms involve a number of free
parameters (e.g., number of clusters, number of hierarchy levels, citation
threshold, etc.). The values of the free parameters are fixed from
external considerations. There is normally a strong dependence of the
clustering results on the values of the free parameters. As a result, 
variation of the parameters yields a too broad set of clusterings ranging
from the trivial clustering (with a single cluster) to the maximally
refined one.

Under the Open Task IV of the KDD Cup 2003, we formulate the question:
Do there exist nontrivial hierarchical graph clusterings
that would be uniquely determined by the graph structure, or, otherwise,
would be weakly dependent in a certain sense on the free parameters of the 
clustering procedure, if the latter are present? Practically, the 
weak dependence can be defined as a weak dependence of the characterizations
like the number of clusters, the number of hierarchy levels, etc.
We answer the question formulated above in the affirmative, and suggest
a new algorithm, \texttt{EqRank} that performs a hierarchical clustering 
solving the problem. The problem is considered for the case
of directed graphs. The algorithm is applied to the \texttt{hep-th}
citation graph. As a result, we obtain a classification scheme
of the subject field presented in \texttt{hep-th}, and indexing
of papers in this classification scheme.

The reminder of this paper is organized as follows. In section 2, we
give an informal explanation of the \texttt{EqRank} algorithm, formalize the intuitive 
presentation, and demonstrate that \texttt{EqRank} is closely related to
the recursive algorithms exemplified by the HITS algorithm 
\cite{kleinberg:authoritative}. We end this section with a crude 
estimate of the time complexity of \texttt{EqRank}. In section 3, we present
the results yielded by \texttt{EqRank} applied to the \texttt{hep-th} citation graph.
Section 4 contains a discussion of the results obtained, and 
lists some unanswered questions.

\section{The Algorithm}

In this section we describe the \texttt{EqRank} algorithm. It can be
applied to any directed graph. Small modifications may be needed
to fine-tune it to a particular setting. The setting that
have been motivating us in the development of \texttt{EqRank}
is the setting of a citation graph. 

\subsection{Informal Explanation}
\label{informal}
We explain the idea behind the \texttt{EqRank} algorithm using the 
terms natural
for a citation graph. Assume that we have learned somehow 
the way to compute the {\it local hub} paper $LH(p)$ for each paper $p$.
$LH(p)$ cites $p$ and is the most representative paper among
the papers developing the ideas of $p$. The existence of the mapping 
$LH$ generates the trajectory $(p, LH(p), LH(LH(p)), ...)$ consisting
of the sequence of papers where every next paper is the local hub of the
previous paper. The trajectory starts at the paper $p$, and ends at the 
paper $RH(p)$, which has no citations. We call the end point of the trajectory
the {\it root hub} of the paper $p$. Let us introduce an equivalence 
relation on the set of papers: 
$p\sim p^\prime \; {\rm if} \; RH(p)=RH(p^\prime)$. The corresponding 
equivalence classes are called the {\it modern themes}. A modern theme 
is formed with
the papers that share a common resulting paper, the root hub, 
which is the paper underscoring the present state of the root theme.
In complete analogy, starting with existence of the {\it local authority}
$LA(p)$ that is the paper cited by $p$, and is the most representative
paper among the papers on which $p$ is based, we determine the partition of the
set of papers into {\it classic themes}. Each paper in a classic theme
has one and the same paper as its {\it root authority}. Frequently, 
a root authority is a seminal paper initiating a new direction of research.
We call simply the {\it themes} the elements of the partition yielded
by intersection of the hub partition and the authority partition. All
the papers of a theme have one and the same root hub and authority papers.
A modern theme considered as a graph whose vertexes are the papers of
the theme, and the links are the links of the citation graph
of the form ($p$, $LH(p)$) is an out-tree \cite{efe:shape} whose root
coincides with the root hub of the theme. Similarly, a classic theme 
is an in-tree whose root coincides with root authority of the theme.

Restricting consideration to a citation graph, it is natural to define the
mapping value $LH(p)$ as  the paper on which
a weight function $W(p, P)$ reaches its maximum:
\[
W(p,LH(p))=\max_P(W(p,P)).
\]
Here $W(p,P)$ is a nonnegative function defining the weight (relevance) of
the link from $P$ to $p$. We can use as $W$ the co-citation
\cite{small:structure}, the bibliographic coupling 
\cite{kessler:bibliographic}, or other link-based measures of proximity
\cite{lu:node}. In the actual experiment we performed over the 
\texttt{hep-th} citation graph, we used as $W$ a linear combination of the
first two of the measures of proximity mentioned above (more
precisely, the reduction of this function onto the links of the
citation graph under consideration).

\subsection{Formal Description of the Algorithm}
\label{formal}
In the informal description of the algorithm, it was implicitly implied that
there is a unique local hub (authority) for each paper, and that
the graph is acyclic. In applications, both conditions are violated.
The following formal description is valid without 
these simplifying assumptions.

Let $G\equiv G(V,E,W)$ be a weighted directed graph, where $V$ is the set of 
the graph vertexes, $E$ is the binary relation on $V$ that
defines the links, $W$ is the nonnegative function on $E$ that defines the
weight of the links. Let $PS(V)$ be the power set, i.e., the
set of all subsets of $V$, and
$FS(V)$ be the final set, i.e., the subset of $V$ 
singled out by the absence of the 
links outgoing from its elements. $SCR$ denotes the strong connectivity
relation on the vertexes of the graph.

\subsubsection{Auxiliary Operations Acting on the Graphs}
\label{aux}
\begin{definition}
\label{o1}
The result of factoring of $G(V,E,W)$,
\[
G/R=G(V/R,E/R,W^\ast)
\] 
is the factor graph of $G$ taken by the 
equivalence relation $R$. Here $V/R$ is the set of the equivalence classes
with respect to $R$, $E/R$ is the binary relation on $V/R$ induced by
$E$, i.e., $X(E/R)Y$ if there exist such 
representatives of these classes $x$ and $y$ that $xEy$.
(We write $xEy$ if $(x,y)\in E$.) 
The function $W^\ast$ is defined as the sum of the weights of all
the links joining the elements belonging to the different equivalence classes.
\end{definition}
\begin{definition}
\label{o2}
The result of inversion of $G(V,E,W)$,
\[
In(G)=G(V,E^{-1},W^\prime)
\]
is the same graph as $G$, but with the directions of the links inverted;
$W^\prime(p,p^\prime)\equiv W(p^\prime,p)$.
\end{definition}
\begin{definition}
\label{o3}
The result of retaining in $G(V,E,W)$ only the maximally weighted links,
\[
Max(G)=G(V,E_{max},W^\prime)
\]
is the graph whose set of links $E_{max}$ is the subset of the maximally
weighted links, $E_{max}\subseteq E$,
\[
(x,y)\in E_{max}\;{\rm if}\; W(x,y)=\max_z(W(x,z)).
\]
$W^\prime$ is the restriction of the function $W$ on $E_{max}$.
\end{definition}
\begin{definition}
\label{root}
Graph $G$ can be transformed to a function 
$Root(G):V\rightarrow PS(V)$,
\[
Root(G)(p)\subseteq FS(V).
\]
This subset is singled out in $FS(V)$ by the property that each of its
points is reachable from $p$ along the links.
\end{definition}

\subsubsection{Equivalence Relations}
We define the following three equivalence relations on $V$, $HubR(G)$,
$AuthR(G)$, and $EqRank(G)$.

$x\sim y$ with respect to the partition $HubR(G)$ if
\[
Root(Max(In(G))/SCR)(x)=Root(Max(In(G))/SCR)(y);
\]
$x\sim y$ with respect to the partition $AuthR(G)$ if
\[
Root(Max(G)/SCR)(x)=Root(Max(G)/SCR)(y);
\]
The desired $EqRank$ partition is defined as follows:
\[
EqRank(G)=HubR(G)\cap AuthR(G).
\]

Some notes are in order. The operation $Max(G)$ (or $Max(In(G))$) keeps
in the graph only the maximal outgoing (or ingoing) links, which link
the papers to their local authorities (hubs). We do not assume
anymore that local authorities and hubs are unique. Because of this,
a classical (modern) theme has as its root not a single paper, but a
set of papers reachable from each paper of the theme along the maximal 
links. Let us comment on the presence of the factoring with respect
to $SCR$ in the above definitions. Without it, going along
the links could be jammed on the cycles of the graph (see subsection
\ref{recursive} for extra motivation of this factoring).

With the above definitions, the algorithm \texttt{Eqrank} 
we suggest for hierarchical clustering is defined as follows.
\begin{itemize}
\item{
The input of \texttt{EqRank} is a directed weighted graph $G$} 
\item{The 
output of the algorithm is the sequence of reduced graphs
$G\equiv G_0, G_1,...$, where $G_i=G_{i-1}/EqRank(G_{i-1})$}
\end{itemize} 
The
sequence terminates when $G_i\sim G_{i-1}$, i.e., the number
of vertexes of $G_i$ coincides with the number of vertexes of 
$G_{i-1}$. 

\subsection{A Recursive Definition and Related Works}
\label{recursive}
In this subsection, we demonstrate that above equivalence relation
$EqRank$ is a natural development of the recursive algorithms
\texttt{PageRank} \cite{page:pagerank}, \texttt{HITS}
\cite{kleinberg:authoritative}, and \texttt{SimRank}
\cite{jeh:simrank}, which became lately
quite popular among the network miners.

In \cite{jeh:simrank} the proximity measure $SimRank$ was introduced for 
relational data. Its definition is based on the simple idea that close
(similar)
objects should be related to close (similar) objects. We use the same kind
of a recursive definition to define an equivalence relation. In this case,
we say that objects are equivalent if they are linked to equivalent
objects. We will demonstrate that the above $EqRank$ equivalence 
relation results from the above recursive definition. (The similarity
between $SimRank$ and $EqRank$ have motivated the name of the latter
equivalence relation.) Below we give the exact definitions. Let 
$G(V,E)$ be a directed acyclic graph, V be its set of vertexes,
$E$ be a binary relation on $V$. The 
binary relation $E$ defines the mapping $Fe: V\rightarrow PS(V)$:
\[
Fe(x) = \{y\in V: xEy\}.
\]
If the above formula yields $Fe(x)=\emptyset$, we set by definition
$Fe(x)=x$. We extend the mapping $Fe$ to a mapping defined on the whole
set of subsets of $V$, $PS(V)$. This is achieved by the formula
\[
Fe(X)=\cup_{x\in X}Fe(x). 
\]

The equivalence relation we want to define should have the property
that $x\sim y$ if $Fe(x)\sim Fe(x)$. This definition falls short to be 
recursive, because it involves the relation between the vertexes and the
relation between the subsets simultaneously. Since $V$ is a subset of
$PS(V)$, and an equivalence relation on a set generates an equivalence 
relation on any subset, the desired equivalence relation can be
recursively defined as follows:
For any $X,Y\in PS(V)$, $X\sim Y$ 
with respect to the partition 
$EqRank^\prime$ if $Fe(X)\sim Fe(Y)$.

The above definitions imply that $x\sim y$ if $Fe^n(x)\sim Fe^n(y)$ 
for any $n$. Evidently, for any acyclic graph, 
$Fe^n(x)=Root(G)(x)$ for $n$ large enough (the $Root(G)$ is
defined in subsection \ref{aux}). Note that $Root(G)(V)=FS(V)$,
and the mapping $Fe$ acts trivially on $FS(V)$.
Thus, there are many solutions to the above recursive
equation for $EqRank^\prime$.
To single out a particular solution, one has to
define an equivalence relation on $FS(V)$. Let us use 
the finest equivalence relation on $FS(V)$:
\[
X\sim Y\; {\rm if} X=Y,
\]
where $X, Y$ are subsets of $FS(V)$.

At this choice we have
\[
EqRank^\prime(Max(G))=AuthR(G),
\]
and 
\[
EqRank^\prime(Max(In(G)))=HubR(G) 
\]
if $G$ is an acyclic graph. Notice that this $EqRank^\prime(G)$ is the finest
equivalence relation satisfying the above recursive equation.

$EqRank$ is expressed in terms of $EqRank^\prime$ with a simple formula:
\[
Eqrank(G) = EqRank^\prime(Max(G))\cap EqRank^\prime(Max(In(G))).
\]

Lastly, we point out that solving recursively the equation for 
$EqRank^\prime$ can be trapped in cycles if they exist. By this
reason, we apply the factoring with respect to the strong connectivity
relation to reduce the problem to the case of acyclic graph.

\subsection{Time Complexity of the Algorithm}
The algorithm consists of the operations \ref{o1}--\ref{root}. Thus,
we have to estimate the time complexity of these operations. Evidently,
$T_{o1}\sim E+V$, where $E$ is the number of links; 
$T_{o2}\sim E$; $T_{o3}\sim E$. Most time consuming is
the operation \ref{root}, because it requires computation
of the transitive closure on the graph. Its time complexity is
$V(E+V)$. 

We point out that we observed linear dependence of the time
complexity on the number of vertexes up to the scale of $10^4$
for the number of vertexes. This may be related to a considerable
simplification of the graph after the application of the
$Max$ operation (see subsection \ref{aux}). About 76\% (91\%)
of the vertexes of the graph $Max(G)$ ($Max(In(G))$) had the 
unit out-degree in our experiments.

\section{Hierarchical Clustering of hep-th Citation Graph}
\label{hepth}
As mentioned above, application of \texttt{EqRank} in a concrete setting
may require a fine-tuning. Here we specify the modification
of \texttt{EqRank} that was applied to the \texttt{hep-th} citation
graph, and present the results obtained.

The graph under consideration consists of 27,240 vertexes, and 342,437 links.
It contains a number of weakly connected components. The largest component
has 26,870 vertexes. The rest of 370 papers fragments into 229 of small
(less than 5 papers) weakly connected components. A consideration of the 
reference lists of the papers from the small components reveals the
reason behind the presence of these small components: most of the papers
cited from the small components do not belong to \texttt{hep-th}, and, 
therefore, escape from the citation graph under consideration. 

The weight of a link was taken to be a linear combination of
the co-citation and bibliographic coupling, 
\begin{equation}
\label{weight}
W(x,y)=a A^T A + (1-a) A A^T,
\end{equation}
where $A$ is the adjacency matrix of the graph. We set $a= 0.9$ The 
closeness of $a$ to unit reflects that we consider co-citations
as a more adequate measure of the importance of a link
(we do not set $a=1$ to avoid degeneracy of the weight function).

Evidently, the clustering of the weakly connected components can be
performed independently for each component. The results we present below
have been obtained by applying \texttt{EqRank} to the largest weakly connected
component.

\subsection{Determination of the Number of Themes}
Applying literally \texttt{EqRank} to \texttt{hep-th} yielded a too refined
clustering. The total number of clusters turned out to be 11,299.
The reason for the existence of such a large number of small clusters
is as follows. The themes are singled out by their root hubs and authorities
in our approach. The root hubs are characterized by the absence of 
links pointing to them (they are recent papers with no papers citing them).
The total number of such papers is too high to keep them all as representatives
of relevant themes. A similar consideration can be applied to root
authorities after inversion of the links. 

To obtain a meaningful reduction of the number of themes, we considered as
``actual'' themes the themes whose number of papers was exceeding
a cutoff value. In the present experiment the cutoff value was taken to
be 20 papers. See below for an analysis of the dependence 
of the classification on the cutoff. For the
above value of the cutoff, the number of actual themes turned out to
be 136. The rest of the themes were glued to the subset of actual themes, each
small theme, to the ``closest'' large theme. The closeness between
the themes $A$ and $B$ was computed as the sum of weights of the links
between the themes regardless of the direction of a link. Ultimately, the
largest theme turned out to contain 3586 papers, and the smallest, 26 papers.

\subsection{Determination of the Number of the Hierarchy Levels}
The themes formed after the first stage clustering of the \texttt{hep-th}
citation graph form the vertexes of the factor graph (see subsection
\ref{aux}). This factor graph was clustered with \texttt{EqRank}. It 
yielded 19 themes of the second hierarchy level. At this level, the 
largest theme contains 15,410 papers, and the smallest, 65. 
One more application of \texttt{EqRank}
yielded the trivial clustering. All the papers merged to a single cluster.

We point out that the cutoff was used only to generate the first level
of the hierarchy. Let us consider the hierarchy that would
appear without the cutoff. Without the cutoff, a star-like graph would appear
at the third level of the hierarchy instead of the trivial graph
consisting of a single vertex. There would be a single super-cluster
of 7228 papers, and a multitude of small themes each of which 
would be connected to the super-cluster either by ingoing or
outgoing links. Further application of \texttt{EqRank} would shrink
the star-like graph by absorbing a number of the small themes into the
super-cluster. Starting from the third hierarchy level, the exponential
reduction of the number of vertexes in the factor graph would switch over to
linear reduction.

Based on the above we claim that \texttt{EqRank} allows computing 
the number of hierarchy levels implied by the structure of a graph. 
For the \texttt{hep-th} citation graph, there are two levels in the hierarchy.

In conclusion, we discuss in more details 
the dependence of the classification 
obtained for \texttt{hep-th} papers on the cutoff,
which is essentially the only parameter involved in the procedure.
The experiments performed have demonstrated that the  
characteristic most stable against changes in the cutoff is the number of
hierarchy levels. It stays invariable as soon as the cutoff exceeds a 
critical value. (This sets an interesting mathematical problem of
understanding the number of hierarchy levels as an invariant characteristic
of a directed graph.) Next in stability is the number of clusters 
on the first hierarchy level. Reduction
of the cutoff simply adds new small clusters without changing
the upper part of the list of clusters. The most involved 
is the dependence on the cutoff of the higher hierarchy levels. Not only 
new small themes may appear, but there
may be also merging and splitting of clusters in the upper part of 
the cluster list at a reduction of the cutoff.

We summarize the above discussion as follows. We obtained a set of 
classifications depending on a single parameter 
$F_{cut}$, the minimal number of papers in a theme of the first level.
Each classification $C$ is a finite sequence of graphs,
\[
C=\{G_0, G_1(F_{cut}),..., G_{n+1}(F_{cut})\}.
\]
The sequence terminates on a trivial graph (more generally, 
on a graph invariant with respect to application
of \texttt{EqRank}). The length of the sequence is characterized by
the number $n$, the number of the levels in the hierarchy. The latter
also depends on $F_{cut}$. The dependence is as follows:
\[
n(F_{cut}) = const\quad{\rm if}\;F_{cut}\geqslant F_{min},
\]
and
\[
n(F_{cut}) = f(F_{cut})\quad{\rm if}\;F_{cut}<F_{min},
\]
where $f$ is a function growing fast at $F_{cut}$ decreasing.
In the case of \texttt{hep-th}, $const=2$, and $F_{min} = 8$.   
The qualitative change of the dependence of the number of 
hierarchy levels on $F_{cut}$ taking place at $F_{min}$
suggests that it is natural to set 
$F_{cut} \geqslant F_{min}$.

\subsection{Theme Dynamics}
A study of time dynamics of the themes is a 
part of our experiment with \texttt{hep-th}.
Specifically, time dependence of the size of the themes at different
levels of the hierarchy was considered. A brief account of the results is
given in Table 1. There is an overall increase of the number of
papers posted to \texttt{hep-th} each year. It is distributed unevenly between the
themes. Analyzing this distribution allowed us to classify the clusters
into four groups depending on the character of the evolution trend.
The trend was computed for the period 1992--2002. The data is presented 
on the plots at\\
\texttt{http://hepstructure.inr.ac.ru/hep-th/Theme\_dyn.htm}.

The first group of the clusters (the ``$+$'' trend clusters) includes 
``growing'' themes. There are 10 themes in this group. The second group 
(the ``$-$'' trend clusters) includes ``fading'' themes. There are 5 themes
of this sort. The third group (the ``0'' trend clusters) includes 2 themes
characterized  by a stable number of the papers appearing per year. The fourth
group (the ``$++$'' trend clusters) includes 2 themes. They are  
``emergent'' themes characterized by explosive growth of the number of papers
appeared in 2002. 

Let us make a comment on the emergent themes (the fourth group). Based
on the appearance of the plots of time dependence we
speculate that if we would cluster the \texttt{hep-th} citation
graph based on the data restricted to the period 1992--2001 (we plan to
make this exercise), the clusters corresponding to the
emergent themes would not appear at all.
In other words, we speculate that the themes of the fourth group were 
born namely in 2002.

\begin{table*}
\centering
\caption{Theme Dynamics}
\begin{tabular}{|c|c|c|c|} \hline
Theme Number& Number of Papers&Theme Label&Trend\\ \hline
1 &15,410&adf/cft correspondence&$+$\\ \hline
2 &4,118 &non-commutative geometry&$+$\\ \hline
3 &908   &tachyon condensation&$+$\\ \hline
4 &858 &stokes theorem&$+$\\ \hline
5 &673 &iib orientifolds&$+$\\ \hline
6 &578 &form factors; ising model&$-$\\ \hline
7 &515 &affine todda&$-$\\ \hline
8 &501 &dilaton gravity&$-$\\ \hline
9 &477 &higher spin&$+$\\ \hline
{\bf 10} &{\bf 457} &{\bf pp-wave background}&$++$\\ \hline
11 &434 &$n=2$ string&$-$\\ \hline
12 &414&renormalization group&$+$\\ \hline
13 &385 &string cosmology&$0$\\ \hline
14 &376 &random matrix&$+$\\ \hline
15 &233 &bethe ansatz&$0$\\ \hline
16 &180 &geometric entropy&$-$\\ \hline
{\bf 17} &{\bf 180} &{\bf rolling tachyon}&{\bf $++$}\\ \hline
18 &108 &gauged supergravity&$+$\\ \hline
19 &65 &taub-nut background&$+$\\ 
\hline\end{tabular}
\end{table*}

\subsection{An Estimate of the Clustering Quality}
\begin{table*}
\centering
\caption{Community Index}
\begin{tabular}{|c|c|c|c|} \hline
Theme Number& Number of Papers&Theme Label&Comm. Index\\ \hline
1 &15,410&adf/cft correspondence&0.95\\ \hline
2 &4,118 &non-commutative geometry&0.76\\ \hline
3 &908   &tachyon condensation&0.78\\ \hline
4 &858 &stokes theorem&0.86\\ \hline
5 &673 &iib orientifolds&0.68\\ \hline
6 &578 &form factors; ising model&0.87\\ \hline
7 &515 &affine todda&0.89\\ \hline
8 &501 &dilaton gravity&0.83\\ \hline
9 &477 &higher spin&0.62\\ \hline
10 &457 & pp-wave background&0.89\\ \hline
11 &434 &$n=2$ string&0.80\\ \hline
12 &414&renormalization group&0.96\\ \hline
13 &385 &string cosmology&0.81\\ \hline
14 &376 &random matrix&0.79\\ \hline
15 &233 &bethe ansatz&0.72\\ \hline
16 &180 &geometric entropy&0.85\\ \hline
17&180&rolling tachyon&0.66\\ \hline
{\bf 18} &{\bf 108} &{\bf gauged supergravity}&{\bf 0.37}\\ \hline
19 &65 &taub-nut background&0.90\\ 
\hline\end{tabular}
\end{table*}

The quality of the classification obtained for \texttt{hep-th}
can be safely estimated only with an analysis performed by experts in
this subject field (see however the Appendix). In this section,
we give a formal estimate based on the analysis of the citation graph itself.
In \cite{flake:efficient}, in the context of web clustering, 
the notion of ``ideal community''
was introduced. It is a subset of vertexes with the following property:
the sum of weights of the inner outgoing links is bigger than the sum of
weights of
the outer outgoing links. Here the inner links join the vertexes of the 
``ideal community'' subset, and the outer links are the links starting at
the subset, and going out of it. In line with this definition, we
computed the so-called ``community index'', which is the ratio 
of the sum of weights of the inner links of a cluster to the
sum of weights of the inner and outer links. If the community index exceeds
0.5, the community is ideal; the larger the community index, the
more ideal is the community. As an overall characteristic, a weighted
mean value of the community index over all the clusters was
computed. Table 2 gives the values of the community index for
the themes of the second level of the hierarchy. As seen, 18 of the 19
themes comply with the formal definition of the ideal community.
The weighted mean value of the community index is 0.88 for the themes 
of the second level. The situation is less satisfactory for the themes of 
the first hierarchy level. About half of the themes of the
first level comply with the definition of the ideal community.
Despite this, the weighted mean value of the community index exceeds 0.5
at this level also. It is 0.58.


\subsection{Theme Representation}
The problem of presenting a theme was not a central one
for this experiment. Despite this, we gave the themes a
number of attributes helping to grasp the content of a theme,
and recognize it. The list of themes with their
attributes is available at
\texttt{http://hepstructure.inr.ac.ru/hep-th/}.
Specifically, the following attributes were
determined for each theme:
\begin{itemize}
\item
{Theme Label. This is a sequence of seven pairs of words 
naming the theme. These sevens were determined by a modification
of {\it Frequent and Predicative Words Method} explained 
in \cite{popescul:automatic}. This method seeks for a word
that is optimally unique and common for a cluster.
The modification is that we sought not for separate words but
for pairs of consecutive words (in this, prepositions and common words
from a stop list were ignored). The body of the analyzed text was
composed from the titles of the papers of the theme.}
\item{
Authority and Hub Papers. For each paper of a cluster, 
a pair of numbers ({\it Authority Number}, {\it Hub Number}) 
was computed. The{\it Authority Number} is
\[ 
\sum_{p^\prime}W(p,p^\prime), 
\]
where the sum runs over the papers 
of the theme for which $p$ is the local authority paper
(see subsection \ref{informal}). The {\it Hub Number} is the same sum
but running over the papers of the theme for which $p$ is the local 
hub (see subsection \ref{informal}). On the above site, we list the 
first 10 papers ordered by decrease of their {\it Authority Number}, and
{\it Hub Number}.}
\item{
List of Main Authors. For each author whose papers are in a cluster,
the pair of numbers 
({\it Author Authority Number}, {\it Author Hub Number}) 
was computed. The {\it Author Authority (Hub) Number} is the sum of
{\it Authority (Hub) Number} of the papers of the author from the cluster.
On the above site, we list the 
first 10 authors ordered by decrease of their 
{\it Author Authority Number}, and
{\it Author Hub Number}.}
\end{itemize}

\section{Conclusions and Outlook}
To summarize, we suggested a new method of hierarchical clustering
of directed graphs. There is a free parameter in the method,
the minimal acceptable number of vertexes in a cluster.
Qualitative features of the dependence of the resulting number of hierarchy
levels on this parameter allows setting it to a certain interval
of natural values. The time complexity of the algorithm is at worst 
quadratic in the number of vertexes in the graph. 

We applied the method to the \texttt{hep-th} citation graph. In this 
application the weight of the links was defined as a linear combination
of the co-citation and bibliographic coupling. The outcome is a
two-level hierarchy of themes present in \texttt{hep-th}.

We list below a number of problems for the future.

It would be interesting to study the dependence of the clustering 
on the weight function. (We assume here that the weight function
is a function of the graph.) We performed a 
preliminary study of the clustering
dependence on the parameter $a$ involved in the weight
function $W$ (see section \ref{hepth}). We observed that for
an interval of values of $a$, factor graph obtained by the clustering
is independent of $a$. However, some of the papers travel from cluster
to cluster when $a$ changes. 

Another interesting issue is the hierarchical 
clustering for random graphs that are popular
among physicists \cite{newman:structure}. In this case, the 
distribution of the papers among the clusters, the number of 
levels in the hierarchy, and other characteristics
of the clustering would be random quantities depending
on the parameters of the random graph model. For example, we ask what is
the difference between the clustering by \texttt{EqRank} of a graph
generated by the classical (Poisson) graph models
and more realistic models of ``small world''.

In conclusion, \texttt{EqRank} opens up interesting possibilities
in hierarchical clustering of directed graphs. Its use for
clustering of the \texttt{hep-th} citation graph 
resulted in a meaningful classification
of the papers from \texttt{hep-th}.



%
\bibliographystyle{abbrv}
\bibliography{eqrank}  
%
%

\appendix
\section{An Imitation of an Expert Estimate}
Seeking for an imitation of an expert estimate of our clustering
we selected arbitrarily a theme of the second level, the ``string
cosmology'' theme. Google helped to find the home page
\texttt{http://www.ba.infn.it/$\sim$gasperin/}
of M. Gasperini, whose name
is among the three top authority authors of the theme.
At his page, M. Gasperini says that his page is devoted to 
``string cosmology''. There is a list of more than 100 papers 
on the subject at the site. We selected the papers from this list
that are present in \texttt{hep-th}, and compared the resulting list
with the list of papers of the ``string cosmology'' theme generated by
\texttt{EqRank}. It turned out that 84\% of the papers selected by
M. Gasperini are also selected by \texttt{EqRank}. 

We did the same for a theme of the first level. It was the 
``two-time physics'' theme. There is a brief review of this theme
on the personal page of Itzhak Bars\\
(\texttt{http://physics.usc.edu/$\sim$bars/twoTph.htm})
who keeps the first position in the list of Authority Authors
of the theme. There is a list of 19 papers in the review. All
of them were selected by \texttt{EqRank}.

These facts support
our opinion that \texttt{EqRank} generates an adequate classification
of \texttt{hep-th}. 

\balancecolumns 
\end{document}